\documentclass[twocolumn,showpacs,preprintnumbers,amsmath,amssymb,superscriptaddress]{revtex4}
\usepackage{amssymb}
\usepackage{graphicx,color}
\usepackage{bm}
\begin{document}

\title{Bulk matter evolution and extraction of jet transport parameter in heavy-ion collisions at RHIC}

\author{Xiao-Fang Chen}

\affiliation{Institute of Particle Physics and Key Laboratory of Quark $\&$ Lepton Physics, Huazhong Normal
University, Wuhan 430079, China}


\affiliation{Institut f\"ur Theoretische Physik, Johann Wolfgang
Goethe-Universit\"at, Max-von-Laue-Str. 1, D-60438 Frankfurt am
Main, Germany}

\author{Carsten Greiner}
\affiliation{Institut f\"ur Theoretische Physik, Johann Wolfgang Goethe-Universit\"at,
Max-von-Laue-Str. 1, D-60438 Frankfurt am Main, Germany}

\author{Enke Wang}

\affiliation{Institute of Particle Physics and Key Laboratory of Quark $\&$ Lepton Physics, Huazhong Normal
University, Wuhan 430079, China}


\author{Xin-Nian Wang}
\affiliation{Nuclear Science Division, MS 70R0319, Lawrence Berkeley
National Laboratory, Berkeley, CA 94720}
\affiliation{Institut f\"ur Theoretische Physik, Johann Wolfgang
Goethe-Universit\"at, Max-von-Laue-Str. 1, D-60438 Frankfurt am
Main, Germany}

\author{Zhe Xu}

\affiliation{Frankfurt Institute for Advanced Studies,
Ruth-Moufang-Str. 1, D-60438 Frankfurt am Main, Germany }

\affiliation{Institut f\"ur Theoretische Physik, Johann Wolfgang
Goethe-Universit\"at, Max-von-Laue-Str. 1, D-60438 Frankfurt am
Main, Germany}

\date{\today}

\begin{abstract}

Within the picture of jet quenching induced by multiple parton
scattering and gluon bremsstrahlung, medium modification of parton
fragmentation functions and therefore the suppression of large
transverse momentum hadron spectra are controlled by both the value
and the space-time profile of the jet transport parameter along the
jet propagation path. Experimental data on single hadron suppression
in high-energy heavy-ion collisions at the RHIC energy are analyzed
within the higher-twist (HT) approach to the medium modified
fragmentation functions and the next-to-leading order (NLO)
perturbative QCD (pQCD) parton model. Assuming that the jet
transport parameter $\hat q$ is proportional to the particle number
density in both QGP and hadronic phase, experimental data on jet
quenching in deeply inelastic scattering (DIS) off nuclear targets
can provide guidance on $\hat q_{h}$ in the hot hadronic matter.
One can then study the dependence of extracted initial value of jet
quenching parameter $\hat q_{0}$ at initial time $\tau_{0}$ on the
bulk medium evolution. Effects of transverse expansion, radial flow,
phase transition and non-equilibrium evolution are examined. The
extracted values are found to vary from $\hat q_{0}\tau_{0}= 0.54$
GeV$^{2}$ in the (1+3)d ideal hydrodynamic model  to 0.96 GeV$^{2}$
in a cascade model, with the main differences coming from the
initial non-equilibrium evolution and the later hadronic evolution.
The overall contribution to jet quenching from the hadronic phase,
about 22-44\%, is found to be significant. Therefore, realistic
description of the early non-equilibrium parton evolution and later
hadronic interaction will be critical for accurate extraction of the
jet transport parameter in the strongly interacting QGP phase in
high-energy heavy-ion collisions.

\end{abstract}

\pacs{12.38.Mh,24.85.+p,25.75.-q}

\maketitle

\section{Introduction}
A large amount of experimental data from the Relativistic Heavy Ion
Collider (RHIC) \cite{rhic1,rhic2,rhic3,rhic4} strongly suggest that
a novel form of matter, a strongly coupled quark gluon plasma
(sQGP), may have been formed in the central region of high-energy
heavy-ion collisions. Among the experimental evidences for the
formation of sQGP are the jet quenching phenomena that include the
strong suppression of hadron spectra with large transverse momentum
in central $Au+Au$ collisions as compared to $p+p$ collisions
\cite{star-suppression,phenix-suppression} and the disappearance of
back-to-back correlations of large transverse momentum
hadrons~\cite{star-dihadron}. These jet quenching patterns observed
at RHIC are in good agreement with the theoretical predictions for
jet quenching~\cite{quenching-1992,theory1,theory2,Wang:1998ww,theory3,theory4}
or suppression of large transverse momentum hadrons as a consequence
of parton energy loss and medium modified parton fragmentation
functions induced by multiple scattering as partons propagate
through the dense medium after their initial production. The parton
energy loss and medium modification of the fragmentation functions
due to multiple parton scattering and induced gluon bremsstrahlung
are controlled by the jet transport parameter \cite{Baier:1996sk},
\begin{equation}
\hat q_R=\rho \int dq_T^2 \frac{d\sigma_R}{dq_T^2} q_T^2.
\label{eq:qhat0}
\end{equation}
or the average squared transverse momentum broadening per unit
length for a jet in color representation $R$, which is also related
to the gluon distribution density of the
medium\cite{Baier:1996sk,CasalderreySolana:2007sw} and therefore
characterizes the medium property as probed by an energetic jet.
Here we consider a picture in which the jet parton interacts with
medium particles or quasi-particles with cross section $\sigma_{R}$
and $\rho$ is the local particle density of the medium. Study of the
jet quenching phenomena therefore can provides important information
on the space-time profile of the jet transport parameter and
consequently properties of the sQGP in heavy-ion collisions.

Extensive phenomenological studies on the suppression of single
hadron spectra \cite{Vitev:2002pf,Wang:2003mm,theory1,Eskola:2004cr,Renk:2006nd} and dihadron
correlations \cite{Owens,zhanghz} at large transverse momentum have
been carried out since the first observation of the strong jet
quenching phenomena. Recent emphasis of the phenomenological studies
has been shifted to systematic analyses of the experimental data with different
jet quenching models and qualitative extraction of the jet
transport parameter
\cite{Majumder:2007ae,Bass:2008rv,Bass:2008ch,Qin:2009gw,Armesto:2009zi}.
In this paper, we will carry out phenomenological analysis of the
suppression of single hadron spectra within the higher-twist (HT)
formalism for medium modification of the parton fragmentation
functions and the next-to-leading order perturbative QCD parton
model \cite{Owens} for initial jet production. We will calculate the
single hadron suppression within three different models for the
dynamical evolution of the bulk matter: (1+1)d Bjorken model, (1+3)d
ideal hydrodynamic model and a parton cascade model. We will study
in particular the effect of collective expansion, transverse flow,
non-equilibrium and most importantly jet quenching in the hadronic phase of
the bulk matter. Since the jet transport
parameter is related to the gluon distribution density of the
bulk medium, one expects it to be different in a QGP and hadronic
matter. We will use the information on jet transport parameter in cold
nuclei extracted from phenomenological studies of deeply inelastic
scattering (DIS) off large nuclei \cite{Deng:2009qb} and extrapolate
to a hot hadronic gas assuming that the gluon distribution in a
nucleon in cold nuclei is the same as in a hot hadronic gas. This
allows us to focus on the values of the jet transport parameter in a
QGP that can be accommodated by the experimental data on single
hadron suppression and  the effect of dynamic evolution of the bulk
medium.

The rest of the paper is organized as follows. In the next section, we will review the energy
loss and modified fragmentation functions in hot nuclear medium within the HT approach and their
dependence on the space-time profile of the jet transport parameter $\hat q_{R}$.  We then
describe the next-to-leading order (NLO)
pQCD parton model for single hadron production in heavy-ion collisions in Sec.~\ref{sec:b}.
In Sec .~\ref{sec:c}, we will introduce three different dynamic evolutions for the bulk medium. The numerical
calculations and phenomenological studies of the experimental data on single hadron suppression
and extraction of the jet transport parameters within each dynamic model are carried out in Sec .~\ref{sec:d}.
Finally, we conclude in Sec.~\ref{sec:e} with a summary.

\section{Energy loss and modified fragmentation functions}\label{sec:a}

\begin{figure}
\centerline{\includegraphics[width=6cm]{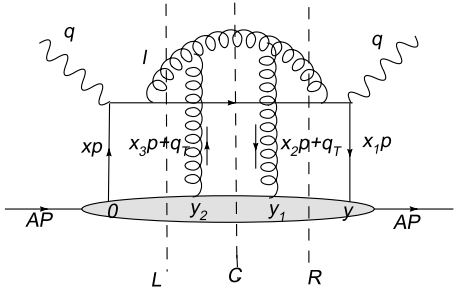}} \caption{A
typical Feymann diagram for quark-gluon re-scattering processes in
DIS with three possible cuts, central(C), left(L), and right(R).}
\label{fig:scatter-in-nucleon}
\end{figure}

Within the generalized factorization of twist-four processes,
one can calculate the nuclear modification of fragmentation functions and the energy loss of a
quark propagating through nuclear matter after it is produced via a hard
process in DIS off a nuclear target \cite{guoxiaofeng,benwei-nuleon}. Within such HT approach,
the medium modification to the quark fragmentation functions in DIS off a nuclear target is caused by
multiple scattering between the struck quark and the nuclear medium on the quark's propagation path
and the induced gluon bremsstrahlung as illustrated in Fig.~(\ref{fig:scatter-in-nucleon}).
The medium modified quark fragmentation function,
\begin{eqnarray}
\tilde{D}_{q}^{h}(z_h,Q^2) &=&
D_{q}^{h}(z_h,Q^2)+\frac{\alpha_s(Q^2)}{2\pi}
\int_0^{Q^2}\frac{d\ell_T^2}{\ell_T^2} \nonumber\\
&&\hspace{-0.7in}\times \int_{z_h}^{1}\frac{dz}{z} \left[ \Delta\gamma_{q\rightarrow qg}(z,x,x_L,\ell_T^2)D_{q}^h(\frac{z_h}{z})\right.
\nonumber\\
&&\hspace{-0.2 in}+ \left. \Delta\gamma_{q\rightarrow
gq}(z,x,x_L,\ell_T^2)D_{g}^h(\frac{z_h}{z}) \right] ,
\label{eq:mo-fragment}
\end{eqnarray}
takes a form very similar to the vacuum bremsstrahlung corrections that leads to the
Dokshitzer-Gribov-Lipatov-Altarelli-Parisi (DGLAP) \cite{dglap} evolution equations in pQCD for fragmentation
functions, except that the medium modified splitting functions,
\begin{eqnarray}
\Delta\gamma_{q\rightarrow qg}(z,x,x_L,\ell_T^2)&=&
[\frac{1+z^2}{(1-z)_{+}}T_{qg}^{A}(x,x_L) \nonumber\\
&&\hspace{-0.7in} +\delta(1-z)\Delta
T_{qg}^{A}(x,x_L)]\frac{2\pi\alpha_sC_A}{\ell_T^2N_cf_q^A(x)};
\label{eq:delta-gamma} \\
\Delta\gamma_{q\rightarrow gq}(z,x,x_L,\ell_T^2)&=&\Delta\gamma_{q
\rightarrow qg}(1-z,x,x_L,\ell_T^2),
 \label{eq:equal}
\end{eqnarray}
depend on the properties of the medium through the twist-four quark-gluon correlations inside the nucleus,
\begin{eqnarray}
T^{A}_{qg}(x,x_L) &=& \int \frac{d^2q_T}{(2\pi)^2}\int
\frac{dy^-}{2\pi}d\xi^{-}dy_2^- d^2\xi_T
e^{i(x+x_{L})p^{+}y^{-}} \nonumber \\
&&\hspace{-1.0in}\times e^{ix_Tp^{+}\xi^{-}-i\vec{q}_T\vec{\xi}_T}
\left\{ (1-e^{-ix_Lp^+y_2^-}) (1-e^{-ix_Lp^+(y^{-}-y_1^-)}) \right.\nonumber \\
&&\hspace{-0.9in}+\frac{1-z}{2} \left[ e^{-ix_Lp^+y_{2}^{-}} (1-e^{-ix_Lp^+(y^{-}-y^-_1)})\right. \nonumber \\
&&\hspace{-0.9in}+\left.\left. e^{-ix_{L}p^{+}(y^{-}-y_{1}^{-})}(1-e^{ix_Lp^{+}y_{2}^{-}}) \right] \right\}\theta(-y_2^-)\theta(y^{-}-y_1^{-})\nonumber \\
&&\hspace{-0.9in} \times\langle A | \bar{\psi}_q(0)\,
\frac{\gamma^+}{2}F_{\sigma}^{\ +}(y_{2}^{-})\,
F^{+\sigma}(y_{1}^-,\xi_T)\,\psi_q(y^-) | A\rangle  ,
\label{eq:qgmatrix}
\end{eqnarray}
where $\xi=y_1 - y_2$, $y$, $y_1$ and $y_2$ are space-time
coordinates associated with the quark and gluon fields as
illustrated in Fig.~\ref{fig:scatter-in-nucleon}. Here we include contributions beyond the helicity
approximation \cite{benwei-nuleon} and consider only the twist-four matrices that are enhanced
by the large nuclear size.  The relative transverse coordinate $\xi_T$ is the Fourier conjugate of the
transverse momentum $q_T$ in the gluon distribution function. The momentum
fraction
\begin{eqnarray}
x_L=\frac{\ell_T^2}{2z(1-z)p^{+}q^{-}}, \label{eq:x_L}
\end{eqnarray}
is the total  (+component) longitudinal momentum transfer from the target
to the propagating quark and radiated gluon with longitudinal momentum $zq^{-}$ and
transverse momentum $\ell_T$.  We impose kinematic
constraints $x_{L}\leq 1$ and $\ell_{T}\leq \min [zE,(1-z)E]$
in the integration in Eq.~(\ref{eq:mo-fragment}).
A similar (+) longitudinal momentum transfer
\begin{eqnarray}
x_T=\frac{q_T^2-2q_T\cdot \ell_T}{2zp^{+}q^{-}}, \label{eq:x_T}
\end{eqnarray}
is also provided by the initial gluon with transverse momentum $q_{T}$.
In the modified splitting function,
\begin{eqnarray}
f_{q}^{A}(x)=\int\frac{dy^-}{2\pi}e^{ix p^+y^-}\langle
A|\bar{\psi}_q(0)\frac{\gamma^+}{2}\psi_q(y^-)|A\rangle
\end{eqnarray}
is the quark distribution function which represents the production rate of the initial quark
carrying $x=Q^2/2p^{+}q^{-}$ (the Bjorken variable) fraction of the nucleon (+) longitudinal
momentum in DIS.  The quark-gluon matrix element,
\begin{eqnarray}
\Delta T_{qg}^{A}(x,x_L) &=& \int_0^{1}dz\frac{1}{1-z}\left[2T_{qg}^{A}(x,x_L)|_{z=1}\right. \nonumber\\
&&\hspace{-0.7in}\left.-(1+z^2)T_{qg}^{A}(x,x_L)\right]
\end{eqnarray}
comes from the virtual correction to the induced gluon bremsstrahlung process.

If we define the quark energy loss as the energy carried by the radiative gluon in the multiple
scattering processes, the total energy loss for a propagating quark in a deeply inelastic scattering (DIS)
off a large nucleus is
\begin{eqnarray}
\frac{\Delta E}{E} = \frac{C_A\alpha_s^{2}}{N_c}\int_0^{Q^2}
\frac{d\ell_T^2}{{\ell_T^4}}
\int_0^1 dz (1+z)^2
\frac{T_{qg}^{A}(x,x_L)}{f_q^A(x)}.
\label{eq:dis0}
\end{eqnarray}

\begin{figure}
\centerline{\includegraphics[width=6cm]{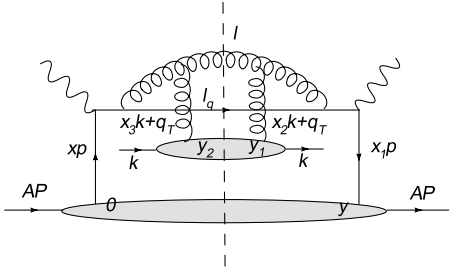}} \caption{Feymann
diagram for induced gluon radiation in hot QGP medium that
contributes the quark energy loss.} \label{fig:scatter-in-QGP}
\end{figure}

One can extend the results for medium modified parton fragmentation
functions in DIS to the case of quark propagation in a hot medium
(either QGP or hot hadronic matter) after it is produced through
initial hard processes before the formation of the QGP
\cite{CasalderreySolana:2007sw,Majumder:2007ae} in high-energy
heavy-ion collisions. In this case, we can assume the quark is
produced through a hard process such as the virtual-photon-nucleus
collisions in Fig.~\ref{fig:scatter-in-QGP} and will interact with
partons in the hot medium which is independent of the initial quark
production process. Therefore, we can replace the nucleus state as a
product of the initial nucleus state and a thermal ensemble of
quasi-particle states in the hot medium,
\begin{equation}
|A\rangle \rightarrow \int\frac{d^3k}{(2\pi)^32k^+}\psi_{k}(y)e^{ik\cdot y}|k\rangle\otimes|A\rangle,
\end{equation}
and consider only the final state interaction between
the produced quark and the bulk medium as illustrated in Fig.~\ref{fig:scatter-in-QGP}.
Here $f(k,y)=|\psi_{k}(y)|^{2}$ is the
local phase-space density of the quasi-particle distribution in the medium. We also
neglect multiple particle correlations inside the hot medium which can be included through
an effective gluon distribution density \cite{CasalderreySolana:2007sw}.
Under these assumptions, the quark-gluon correlation function in the HT approach to
multiple scattering will take a factorized form,
\begin{eqnarray}
\frac{T^{A}_{qg}(x,x_L)}{f_q^A(x)} &=&\frac{N_{c}^{2}-1}{4\pi\alpha_sC_{R}}\frac{1+z}{2} \int dy^{-}
2 \sin^{2}\left[\frac{y^{-}\ell_{T}^{2}}{4Ez(1-z)}\right]
\nonumber \\
&&\hspace{0.0 in}\times\left[\hat{q}_R(E,x_L,y)+c(x,x_{L}) \hat{q}_R(E,0,y)\right] \, , \label{eq:corr2}
\end{eqnarray}
where $c(x,x_{L})=f_{q}^{A}(x+x_{L})/f_{q}^{A}(x)$ and $\hat{q}_R$ is the generalized jet transport parameter,
\begin{eqnarray}
\hat q_R(E,x_L,y)&=&\frac{4\pi^2\alpha_sC_{R}}{N_{c}^{2}-1}
\int\frac{d^3k}{(2\pi)^3} f(k,y)\nonumber \\
&&\hspace{-0.2in}\times\int \frac{d^2q_T}{(2\pi)^2}
\phi_k(x_T+x_L,q_T), \label{eq:qhat-phi}
\end{eqnarray}
which depends on the fractional (+) longitudinal momentum transfer
$x_L$ from medium gluons \cite{note}. Here $\phi_k(x,q_T)$ is the transverse momentum dependent gluon
distribution function from the quasi-particle in the medium,
\begin{eqnarray}
\phi_k(x,q_T) &=& \int \frac{d\xi^-}{2\pi k^+} d^2\xi_T\,
e^{ix k^+\xi^- - i{\bf q}_T\cdot {\bf \xi}_T} \nonumber \\
&&\hspace{-0.1in}\times\langle k| F_{\sigma
}^+(0)F^{+\sigma}(\xi^-,{\bf \xi}_T)|k\rangle\, .
\end{eqnarray}

The two terms in the square brackets of Eq.~(\ref{eq:corr2}) correspond  to two
different processes in the multiple scattering. The first term involves (+) longitudinal momentum
transfer $x_{L}$ between the propagating parton and the medium due to final gluon production.
It contains what is normally defined as pure elastic energy
loss~\cite{elastic}. The second term that is proportional to the normal (or special) transport parameter
$\hat{q}_{R}(E,y)=\hat{q}_{R}(E,x_L=0,y)$ corresponds to pure
radiative processes. In this paper, we assume $x \gg x_L, x_{T}$ and only focus on
the contribution of radiative energy loss.  Therefore, $c(x,x_{L})\approx 1$,
$\hat{q}_{R}(E,x_L,y)\approx \hat{q}_{R}(E,0,y) \equiv \hat{q}_{R}(E,y)$. Given the space-time
profile of the jet transport parameter $\hat{q}_{R}(E,y)$, one should be able to calculate the modified
fragmentation function according to Eq.~(\ref{eq:mo-fragment}). The corresponding quark energy loss
in Eq.~(\ref{eq:dis0}) can be expressed as
\begin{eqnarray}
\frac{\Delta E}{E} &=& \frac{N_{c}\alpha_s}{\pi} \int dy^-dz
{d\ell_\perp^2}
\frac{(1+z)^3}{\ell_T^4} \nonumber \\
&& \hspace{-0.5in}\times \hat q_R(E,y)
\sin^2\left[\frac{y^-l_T^2}{4Ez(1-z)}\right], \label{eq:de-twist}
\end{eqnarray}
in terms of the jet transport parameter. The jet transport parameter for a gluon is $9/4$
times of a quark and therefore the radiative energy loss of a gluon jet is also $9/4$ times
larger than that of a quark jet. In our following calculations of the hadron spectra in heavy-ion
collisions, we will use the same formalism for both quark and gluon modified fragmentation
functions but with jet transport parameters that differ by a factor of 9/4 for quark and gluon jets.

\section{Single hadron spectra within NLO pQCD parton model}\label{sec:b}

In this paper, we employ the NLO pQCD parton model for the initial
jet production spectra which has been shown to work well for large
$p_T$ hadron production in high energy nucleon-nucleon
collisions~\cite{owens1987}. In a factorized form, the inclusive
particle production cross section in $p+p$ collisions can be
calculated as a convolution of parton distribution functions inside
the proton, elementary parton-parton scattering cross sections and
parton fragmentation functions,
\begin{eqnarray}
\frac{d\sigma^h_{pp}}{dyd^2p_T}&=&\sum_{abcd}\int
dx_adx_bf_{a/p}(x_a,\mu^2)f_{b/p}(x_b,\mu^2) \nonumber \\
&&\hspace{-0.1in}\times\frac{d\sigma}{d\hat{t}}(ab\rightarrow cd)
\frac{D_{h/c}^0(z_{c},\mu^2)}{\pi z_{c}}+\mathcal {O}(\alpha_s^3),
\label{eq:pp}
\end{eqnarray}
where $d\sigma/d\hat{t}(ab\rightarrow cd)$ are elementary parton
scattering cross sections at leading order (LO) $\alpha_s^2$. The
next-to-leading order (NLO) contributions including
$2\rightarrow3$ tree level contributions and 1-loop virtual corrections
to $2\rightarrow2$ tree processes. We refer to
Ref.~\cite{Owens} for a detailed description.
The proton parton distribution functions (PDF) $f_{a/p}(x_a,\mu^2)$ are
given by the CTEQ6M parametrization \cite{distribution} where $x_a$ is the fractional
momentum carried by the beam partons.  The fragmentation function (FF) $D^0_{h/c}(z_c,\mu^2)$
of parton $c$ into hadron $h$, with $z_c$ the momentum fraction of a parton jet carried by a produced
hadron are given by the  updated AKK parametrization~\cite{akk08}, which is recently improved
with new input from the RHIC data, especially for $\pi^\pm$, $K^\pm$,
$p/\bar{p}$, $K_S^0$ and $\Lambda/\bar{\Lambda}$ particles.

We use a NLO Monte Carlo based program~\cite{Owens} to calculate the
single hadron spectra in our study. In this NLO program,  the
factorization scale and the renormalization scale are chosen to be the
same (denoted as $\mu$) and are all proportional to the transverse momentum of the final hadron $p_T$.
As pointed out in Ref.~\cite{zhanghz}, the calculated single inclusive pion spectra
within the NLO pQCD parton model for $p+p$ collisions agree well with the
experimental data at the RHIC energy using the scale in the range
$\mu=0.9 \sim 1.5 p_T$. We find that the calculated $\pi^{0}$ spectra in
$p+p$ collisions with the scale $\mu=1.2p_T$ can fit RHIC data
very well and will use the same scale in $A+A$ collisions in our
following calculations.

We further assume that the factorized form for large transverse momentum single
hadron spectra in NLO pQCD parton model can be applied to heavy-ion collisions,
\begin{eqnarray}
\frac{1}{N_{\rm bin}^{AB}(b)}\frac{d\sigma_{AB}^h}{dyd^2p_T} &=&\sum_{abcd}\int
dx_adx_b f_{a/A}(x_a,\mu^2)f_{b/B}(x_b,\mu^2) \nonumber \\
&&\hspace{-0.5in}\times \frac{d\sigma}{d\hat{t}}(ab\rightarrow
cd)\frac{\langle \tilde{D}_{c}^{h}(z_{h},Q^2,E,b)\rangle}{\pi z_{c}}+\mathcal {O}(\alpha_s^3), \nonumber \\
\label{eq:AA}
\end{eqnarray}
where $N_{\rm bin}^{AB}(b)$ is the number of binary nucleon-nucleon collisions at a fixed impact $b$ of
$A+B$ nuclear collisions, $\langle \tilde{D}_{c}^{h}(z_{h},Q^2,E,b)\rangle$ is the medium modified
parton fragmentation function averaged over the initial production position and jet propagation direction,
and  $f_{a/A}(x_{a},\mu^{2})$ is the effective parton distributions per nucleon inside a nucleus. For
large transverse momentum hadron production, the effect of nuclear modification of the parton
distributions is small. We will neglect such nuclear effect in our study here.

According to the definition of the jet transport parameter in
Eq.~(\ref{eq:qhat-phi}), it should be proportional to the local
particle density at $\vec r_{j}=\vec r +(\tau-\tau_{0})\vec n$ in
either QGP or hadronic phase of the evolving bulk medium
along the path of a propagating jet  which is a straight line in the
eikonal approximation. Here $\tau_{0}$ is the formation time of the
medium and the direction of jet propagation is defined by its
azimuthal angle $\phi$ in the transverse plane. The quark-gluon
correlation in Eq.~(\ref{eq:corr2}) that enters the modified
fragmentation function in Eq.~(\ref{eq:mo-fragment}) will involve an
integration over the duration of the jet propagation and depend on
the initial production position $\vec r$ and propagation direction
$\vec n$. In heavy-ion collisions at a fixed impact parameter $\vec
b$, the initial jet production cross section is proportional to the
overlap function of two colliding nuclei $\int d^{2}r
t_{A}(r)t_{B}(|\vec b -\vec r|)$. One therefore has to average over
the initial jet production position and propagation direction to
obtain the effective medium modified parton fragmentation function,

\begin{eqnarray}
\langle \tilde D_{a}^{h}(z_h,Q^2,E,b)
\rangle &=& \frac{1}{\int
d^2{r}t_{A}(|\vec{r}|)t_{B}(|\vec{b}-\vec{r}|)} \nonumber \\
&&\hspace{-1.0in}\times \int\frac{d\phi}{2\pi}d^2{r}
t_{A}(|\vec{r}|)t_{B}(|\vec{b}-\vec{r}|)\tilde{D}_{a}^{h}(z_h,Q^2,E,r,\phi,b). \nonumber \\
\label{eq:frag}
\end{eqnarray}

With a given space-time profile of the jet transport parameter, one can therefore use the above
effective medium modified fragmentation functions to calculate the large transverse
momentum hadron spectra and the the suppression factor (or nuclear modification factor),
\begin{eqnarray}
R_{AB}(b)=\frac{d\sigma_{AB}^h/dyd^2p_T}{N_{bin}^{AB}(b)
d\sigma_{pp}^h/dyd^2p_T}, \label{eq:rab}
\end{eqnarray}
for characterization of the effect of jet quenching on hadron spectra in heavy-ion collisions \cite{modification-factor}.
Phenomenological study of the experimental data on the above hadron suppression factor will
in turn provide us constraints on the space-time profile of the jet transport parameter $\hat q_{R}(E,y)$.

The computation of the effective medium-modified jet fragmentation function in Eq.~(\ref{eq:frag}) involves
a 6-dimensional integration which is in addition to the Monte Carlo integration in the NLO pQCD calculation of
the single hadron spectra~\cite{Owens}. Numerically, we will compute and tabulate the medium modified fragmentation functions
as functions of the initial jet energy $E$, fractional momentum $z$ and the factorization scale $Q^{2}$ for  fixed value of
impact parameter $b$. We will then
use numerical interpolation of the table for the medium modified fragmentation functions in the NLO pQCD calculation of the single
hadron spectra.

\section{Bulk matter evolution}\label{sec:c}

Recent phenomenological studies
\cite{CasalderreySolana:2007sw,Majumder:2007ae,Bass:2008rv,Bass:2008ch,Qin:2009gw}
have focused on the initial values of the jet transport parameters
as constrained by the experimental data. We study in this paper
the uncertainty of the extracted initial jet transport parameter due
to different models of the dynamical evolution of the bulk
medium and in particular the effect of transverse expansion, flow
effect, non-equilibrium and phase transition by considering three
different models for the bulk evolution. Assuming fast
thermalization of the partonic matter in the initial stage of
high-energy heavy-ion collisions,  we will consider first (1+1)d Bjorken
\cite{Bjorken:1982qr}  and (1+3)d ~\cite{hirano1, hirano2} ideal
hydrodynamic  evolution of the bulk matter for the calculation of
medium modified parton fragmentation functions. Inclusion of shear
and bulk viscosity in the viscous hydrodynamics can affect
significantly the elliptic flow or azimuthal asymmetry of the final
hadron spectra at freeze-out. Their effect on the space-time profile
of the bulk medium such as entropy density, however, is much smaller
(on the order of a few percent)
\cite{Teaney:2003kp,Romatschke:2007mq,Song:2007ux,Dusling:2007gi}
which we will neglect here in the hydrodynamic evolution of the bulk
matter for the study of jet quenching. To consider non-equilibrium
evolution of the bulk matter and its effect on jet quenching, we
will also calculate medium modified hadron spectra with the
space-time profile of parton density as given by a parton cascade
model~\cite{Xu:2004mz,Xu:2007aa}.

\subsection{ (1+1)d Bjorken Expansion}\label{sec:c-a}

In a (1+1)d Bjorken model \cite{Bjorken:1982qr} for the bulk
evolution, the system is assumed to undergo a longitudinally
boost-invariant expansion with the (1+1)d ideal hydrodynamic equation,
\begin{equation}
\frac{d\epsilon}{d\tau}=-\frac{\epsilon+P}{\tau},
\end{equation}
where $\epsilon$ is the energy density, $P$ is the pressure and
$\tau=\sqrt{t^{2}-z^{2}}$ is invariant time. For a massless ideal
gas equation of state (EOS), the solution of the above ideal
hydrodynamic equation leads to a time evolution of the entropy
density,
\begin{equation}
s =s_{0}\frac{\tau_{0}}{\tau},
\end{equation}
with $s_{0}$ the entropy density at the initial time $\tau_{0}$.
Since the jet transport parameter $\hat q_{R}$ is directly
proportional to gluon distribution density and therefore
proportional to the particle number or entropy density, we assume
that it will have a similar time dependence due to longitudinal
expansion. In the (1+1)d Bjorken model, we further neglects the
transverse expansion and phase transition. Therefore the transverse
profile of jet transport parameter will be given by the initial
transverse profile of the parton density which we assume to be
proportional
 to the transverse density of participant nucleons
according to the wounded nucleon model of nucleus-nucleus collision~\cite{Bialas:1976ed},
\begin{eqnarray}
\hat{q}= \hat{q}_0
\frac{n_{\rm{part}}(\vec{b},\vec{r})}{n_{\rm{part}}(\vec{0},0)}
\frac{\tau_0}{\tau}\,,
\label{q-hat-npart}
\end{eqnarray}
where $n_{\rm{part}}(\vec{b},\vec{r})$ is the transverse density of participant nucleons in nucleus-nucleus
collisions with impact parameter $b$,
\begin{eqnarray}
n_{\rm{part}}(\vec{b},\vec{r})&=& t_A(|\vec{r}|)
\left(1-e^{-\sigma_{NN}t_{A}(|\vec{b}-\vec{r}|)}\right ) \nonumber\\
&& +\ t_A(|\vec{b}-\vec{r}|)\left(1-e^{-\sigma_{NN}t_{A}(|\vec{r}|)}\right)\,,
\label{npart}
\end{eqnarray}
and $\sigma_{NN}$ is the nucleon-nucleon total inelastic cross section, which is set to be
$\sigma_{NN}=42$ mb at the RHIC energy.  Here $t_A(\vec{r})$ is the nuclear thickness function,
\begin{equation}
t_A(\vec{r})=\int_{-\infty}^{\infty} dz\, \rho_A(\vec{r}, z)\,,
\end{equation}
and $\rho_A(\vec{r}, z)$ is the single nucleon density in the
nucleus normalized to $\int d^{2}r t_{A}(\vec r)=A$. The total
number of participant nucleons in nucleus-nucleus collisions at
impact-parameter $b$ is then $N_{\rm part}(b)=\int d^{2}r n_{\rm
part}(\vec b,\vec r)$. In the space-time profile of the jet
transport parameter in Eq.~(\ref{q-hat-npart}), $\hat q_{0}$ is
defined as the jet transport parameter at the center of the bulk
medium in the most central nucleus-nucleus collisions ($b=0$).

We will consider two different nuclear distributions for
$\rho_A(\vec{r},z)$ in the (1+1)d Bjorken model. One is the
Woods-Saxon distribution:
\begin{equation}
\label{wsf}
\rho_A(\vec{r},z)=\frac{n_0}{1+e^{\frac{\sqrt{|\vec{r}|^2+z^2}-R_A}{d}}}\,,
\end{equation}
where $d=0.662$ fm, $R_A=1.26 A^{1/3}$ fm, and $n_0$ is the
normalization factor given by $\int d^2r dz \,
\rho_A(\vec{r},z)=A$~\cite{Schwierz:2007ve}. We also consider
the simple hard-sphere distribution,
\begin{equation}
\label{hsf} \rho_A(\vec{r},z)=\frac{3A}{4\pi
R_A^3}\theta(R_{A}-\sqrt{|\vec{r}|^{2}+z^{2}}),
\end{equation}
with $R_{A}=1.12A^{1/3}$ for comparison.

In the present work we focus on jet quenching in the $10\%$
most central collisions. By comparing the averaged number of
participating nucleons,
\begin{eqnarray}
\label{av_npart} && \langle N_{\rm part} \rangle = \frac{\int db \,
b N_{\rm part}(b)} {\int db\, b},
\end{eqnarray}
which is given by experiments for the most
$10\%$  central collisions~\cite{Adams:2004bi}, we determine the
averaged value of impact-parameter $b=3.15$ fm for Woods-Saxon
nuclear distribution and $b=2.2$ fm for the hard-sphere
distribution.

\subsection{(1+3)d Ideal Hydrodynamical Expansion}\label{sec:c-b}

To take into account of both longitudinal and transverse expansion in the ideal hydrodynamic
description of the bulk matter evolution, we use the output of calculations
by Hirano {\it et al.}~\cite{hirano1,hirano2} for Au+Au collision at the
RHIC energy with Glauber collisions geometry and the average impact parameter $b=3.2$ fm, corresponding
to $10\%$ most central events. The initial condition for the (1+3)d ideal hydrodynamic equations is fixed such that
the final bulk hadron spectra from the RHIC experiments are reproduced \cite{hirano1,hirano2}.
We use the code provided by Hirano \cite{hirano-p} that interpolates the
numerical data from the hydrodynamic solution on a space-time grid for energy density, temperature,
flow velocity and fraction of QGP phase.

For our purpose, we have to infer parton and hadron number density
from the energy density or temperature provided by the hydrodynamic
solution, which uses a Bag model for the equation of state (EOS)
\cite{Nonaka:2000ek}. In this Bag model, the parton number and energy
density in the QGP phase are,
\begin{eqnarray}
\label{rhoqgp}
\rho^{QGP}&=&\left ( 16  \frac{\zeta{(3)}}{\pi^2}+
36 \frac{3\zeta{(3)}}{4\pi^2}\right ) T^3 \equiv a_1 T^ 3, \\
\epsilon^{QGP}&=&\left (16 \frac{\pi^2}{30}+36 \frac{7\pi^2}{240}\right )
T^4 +B\equiv b_1 T^4 +B, \label{eq:eos2}
\end{eqnarray}
for three quark flavors. The Bag constant $B=(247 \mbox{ Mev})^4$ gives rise to a first-order
phase transition from the QGP  to a hadron resonance gas at the critical temperature $T_c=170$ MeV.

In our calculation of medium modified parton fragmentation functions
we consider jet-medium interaction in both partonic and hadronic
phase throughout the evolution of the bulk matter. Neglecting hadron
correlation in the medium, the jet transport parameter will be
proportional to hadron density and the soft gluon distribution
within each hadron. There have been several phenomenological studies
\cite{Wang:2002ri,Majumder:2004pt,Majumder:2009zu} of jet quenching in deeply inelastic scattering
off large cold nucleus as measured by the HERMES
\cite{Airapetian:2007vu} experiment. Within the same high-twist
approach, the jet transport parameter is found \cite{Deng:2009qb} to
be $\hat q_{N}\approx 0.02$ GeV$^{2}$/fm at the center of the cold
nucleonic matter in a large nucleus. In the (1+3)d ideal
hydrodynamic model, the hadronic phase of the medium evolution is
considered as a hadron resonance gas, in which the jet
transport parameter can be approximate as,
\begin{equation}
\hat q_{h}=\frac{\hat q_{N}}{\rho_{N}}\left[ \frac{2}{3}\sum_{M}\rho_{M}(T)+\sum_{B}\rho_{B}(T)\right]
\label{eq:qhath}
\end{equation}
where $\rho_{N}=n_{0}\approx 0.17$ fm$^{-3}$ is the nucleon density in the center of a large nucleus and the
factor $2/3$ accounts for the ratio of constituent quark numbers in mesons and baryons.  The
hadron density at a given temperature $T$ and zero chemical potential is
\begin{eqnarray}
\sum_{h}\rho_{h}(T)=\frac{T^{3}}{2\pi^{2}}
\sum_{h}\left(\frac{m_{h}}{T}\right)^{2}
\sum_{n=1}^{\infty}\frac{\eta_{h}^{n+1}}{n}K_{2}\left(n\frac{m_{h}}{T}\right),
\end{eqnarray}
where $\eta_{h}=\pm$ for meson (M)/baryon (B). In the paper, we will include all hadron resonances with
mass below 1 GeV. Other hadron resonance gas models normally include hadrons with mass up to 2 GeV which
will only make some differences to our assumed $\hat q_{h}$ close to the phase transition temperature.

\begin{figure}
\centerline{\includegraphics[width=9cm]{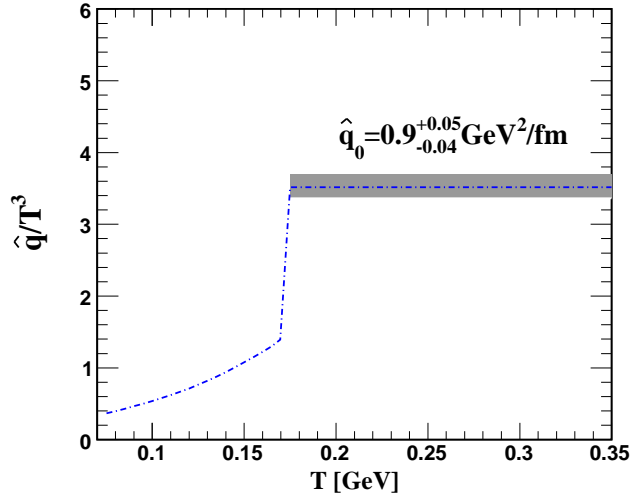}}
 \caption{The temperature dependence of $\hat q/T^{3}$.}
\label{fig:qhat-t3}
\end{figure}

With the above approximation for hadron density and jet transport
parameter in the hadron phase of bulk medium evolution, the jet transport parameter throughout
the evolution of the medium can be expressed as,
\begin{equation}
\label{q-hat-qgph}
\hat{q} (\tau,r)= \hat{q}_0\frac{\rho^{QGP}(\tau,r)}{\rho^{QGP}(\tau_{0},0)}  (1-f) + \hat q_{h}(\tau,r) f \,,
\end{equation}
where $f(\tau,r)$ is the fraction of the hadronic phase at any given
time and local position and $\hat q_{0}$ denotes the jet transport
parameter at the center of the bulk medium in the QGP phase at the
initial time $\tau_{0}$.  To illustrate the behavior of the jet
transport parameter in dense medium, we plot $\hat q/T^{3}$ in
Fig.~\ref{fig:qhat-t3} as a function of the temperature $T$
according to the above assumptions, where we have used the initial
value of $\hat q_{0}=0.9^{+0.05}_{-0.04}  $ GeV$^{2}$/fm, as
extracted by phenomenological study of experimental data later in
this paper, at an initial temperature $T_{0}=$ 369 MeV in the most
central $Au+Au$ collisions at RHIC. Apparently, we have neglected
any additional temperature dependence beyond that in
Eq.~(\ref{q-hat-qgph}). Perturbative calculations
\cite{CasalderreySolana:2007sw,Wang:2000uj} in finite temperature
QCD with resummed hard-thermal loops give rise to a logarithmic
temperature dependence of $\hat q/T^{3}$ for a fixed value of the
strong coupling constant $\alpha_{s}$. The same perturbative calculations also lead
to somewhat weak jet-energy dependence of the jet transport parameter which
we will neglect in this phenomenological study. Therefore, one can consider the exacted
jet transport parameter as an averaged value over the range of jet energies in the
experimental data.

In order to study the effect of phase transition in the (1+3)d hydrodynamic evolution of the
bulk matter in jet quenching, we set $f=0$ in Eq.~(\ref{q-hat-qgph}) and determine the parton density
$\rho= a_{1}[(\epsilon -B)/b_{1}]^{3/4}$ according to Eq.~(\ref{rhoqgp}) throughout the bulk evolution,
with $\epsilon$ given by the (1+3)d hydrodynamic evolution at given space and time. Calculations with
this prescription of space-time profile of the jet transport parameter will be labelled as no phase
transition in this paper. We can also set $\hat{q}_0^h=0$ to quantify the hadronic contribution to jet quenching.

The jet transport parameter $\hat{q}$ in Eq.~(\ref{q-hat-qgph}) is
given in the frame where the medium is at rest. To take into account
of the radial flow, one can simply make the following substitute
\cite{flow1,flow2}
\begin{equation}
\label{q-hat-flow}
\hat{q} \to \hat{q} \frac{p^\mu u_\mu}{p_0} \,,
\end{equation}
where $p^\mu$ is the four momentum of the jet and $u^\mu$ is the
four flow velocity in the collision frame, which is provided by the solution of the (1+3)d
hydrodynamical equations \cite{hirano1,hirano2}. Such flow dependence effectively
decreases (increases) the parton energy loss when the jet propagates
along the same (opposite) direction as the flow.

\subsection{(1+3)d Parton Cascade}\label{sec:c-c}

To study the effect of non-equilibrium evolution of the bulk matter on jet quenching in this paper, we
also consider a parton cascade model. The Boltzmann Approach of MultiParton Scatterings (BAMPS)
model \cite{Xu:2004mz,Xu:2007aa} solves the Boltzmann equation for on-shell gluons with pQCD based
interactions, which include elastic scattering, bremsstrahlung and its back reaction. The thermal equilibration
process of gluons has been thoroughly investigated within this parton transport model \cite{Xu:2007aa,El:2007vg}.
BAMPS can in principle describe collective flow behavior \cite{Xu:2007jv,Xu:2008av,Bouras:2009nn} and
the nuclear modification factor $R_{AA}$ of jet quenching \cite{Fochler:2008ts} within a common framework.

In this work we use the output from BAMPS \cite{Xu:2008av}
model for central $Au+Au$ collisions with $b=3.4$ fm at the RHIC
energy. Initial gluons or minijets are produced by binary
nucleon-nucleon collisions with a Glauber geometry. The Lorentz
contracted nuclei have a longitudinal extent of $0.2$ fm. Thus, most
minijets are produced at $t=0.1$ fm/c when two nuclei fully overlap.
The strong QCD coupling constant is set to be a constant,
$\alpha_s=0.3$.  Gluons freeze out when the local energy density
goes below a critical value, which is chosen as $\epsilon_c=0.6
\mbox{ GeV/}\mbox{fm}^3$. This corresponds to a critical temperature
of $T_c=175$ MeV in a fully equilibrated gluon gas. In the present
implementation of BAMPS there are no hadronization and hadron
cascade. Gluons propagate freely (free-streaming) after freeze-out.
During this stage we regard one gluon as one hadron according to a
parton-hadron duality picture. With these setups the final
transverse energy and the elliptic flow $v_2$ from BAMPS
calculations agree with the experimental data at RHIC
\cite{Xu:2007jv}.

We use Eqs.~(\ref{q-hat-qgph}) and (\ref{q-hat-flow}) to evaluate the jet transport parameter
in the evolving bulk medium as described by the BAMPS model. The local gluon density
and flow velocity are obtained by integrating the local gluon phase space distribution given
by BAMPS. The hadron fraction $f$ is either $0$ before gluon freeze-out or $1$ after gluon freeze-out
as a simplification for the hadronization process. By setting $f=0$ all the time, one effectively neglects the
effect of phase transition and the difference in the jet transport parameter in QGP and hadronic  phase.
The hadron density will be estimated from a free-streaming hadron gas according to an assumption of
hadron-parton duality.

\begin{figure*}[t]
\begin{center}
\includegraphics[width=155mm]{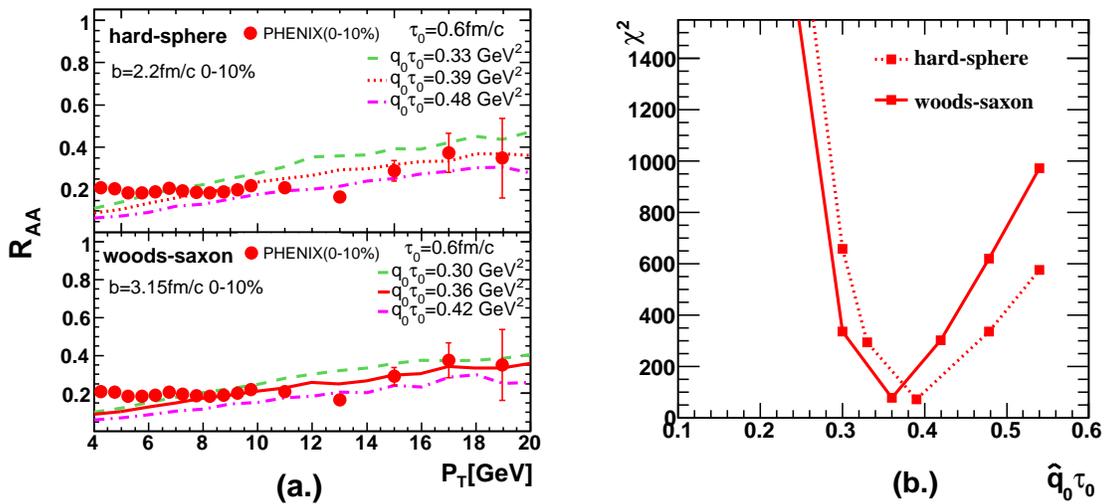}
\caption{(color online) (a) Nuclear modification factor at
midrapidity for the most $10\%$ central Au+Au collisions at the maximum
RHIC energy $\sqrt{s}=200$ GeV/n. The symbols are the PHENIX data taken from
Ref.~\cite{raa}. The curves are NLO pQCD calculations with three different values of
$\hat{q}_0 \tau_0$ in the (1+1)d Bjorken
model with the Hard-Sphere (upper) and Woods-Saxon (lower) distribution function.
(b) The corresponding $\chi^2$ of the fit as a function of $\hat{q}_0\tau_0$.} \label{fig:raa}
\end{center}
\end{figure*}

\section{Numerical results and comparisons}\label{sec:d}

With the three different models for the bulk matter evolution and
the prescriptions for the evaluation of the jet transport parameter
in the evolving medium as described in the last section, we can
conduct a systematic analysis of the experimental data on
suppression of high $p_{T}$ hadron spectra within the higher-twist
approach to the medium modified jet fragmentation functions. Our aim
in this work is to study the theoretical uncertainties associated
with the dynamical evolution of the bulk medium and effects of jet hadron
interaction in the hadronic medium on jet quenching. Since we assume that the jet
transport parameter in the hadronic phase can be estimated from the
value as determined in the cold nuclear matter in the DIS off
nuclear targets [Eq.~(\ref{eq:qhath})], our specific problem in this
case is to determine the value of the jet transport parameter $\hat
q_{0}$ at the center of the overlapped region in the central $Au+Au$
collisions in the QGP phase at the initial time $\tau_{0}$ [see
Eqs.~(\ref{q-hat-npart}) and (\ref{q-hat-qgph})].

Shown in Fig.~\ref{fig:raa}(a) are comparisons between the PHENIX
experimental data on $R_{AA}$ for the neutral pion spectra in the
$10\%$ most central $Au+Au$ collisions \cite{raa} at the highest
RHIC energy $\sqrt{s}=200$ GeV/$n$ and our NLO pQCD calculations
with the medium modified fragmentation functions as given in the
higher-twist approach, for three different values of $\hat{q}_0$ at
the initial time $\tau_0=0.6$ fm/$c$, using (1+1)d Bjorken model of
ideal hydrodynamical evolution without a phase transition. The upper
panel of Fig.~\ref{fig:raa} (a) shows the results with the
hard-sphere nuclear geometry while  the lower panel is for
Woods-Saxon nuclear distribution. The difference between the
calculated suppression factors with hard-sphere and Woods-Saxon
nuclear distributions is about 10 percent, due to the small difference in
the thickness and overlap functions with these two nuclear distributions.

In the remainder of this paper, we will discuss the determination of initial jet transport parameter and its
uncertainties associated with the modeling of the bulk medium evolution.
We will carry out systematical $\chi^{2}$ fits to the experimental data on the suppression factor
$R_{AA}$ of neutral pion spectra from the PHENIX experiment \cite{raa}
with our NLO pQCD calculations that include jet quenching in
an evolving bulk medium. We will follow Ref.~\cite{Adare:2008cg} to carry out a modified $\chi^{2}$ analysis
and obtain the best fit parameters $\hat{q}_0 \tau_0$ for each model of bulk
medium evolution by minimizing
\begin{equation}
\chi^2=\sum_{i=1}^{N}\frac{[y_i+\epsilon_b\sigma_{b_i}+\epsilon_cy_i\sigma_c-y_i^{th}]^2}
{\tilde{\sigma}_i^2}+\epsilon_b^2+\epsilon_c^2\,,
\end{equation}
\begin{equation}
\tilde{\sigma}_i=\sigma_i(\frac{y_i+\epsilon_b\sigma_{b_i}+\epsilon_cy_i\sigma_c}{y_i})\,,
\end{equation}
of the fits with respect to two parameters $\epsilon_b$ and
$\epsilon_c$, where $y_i$ is the central value of experimental data
with three types of uncertainties, $\sigma_i$ (statistical plus
uncorrelated systematics), $\sigma_{b_i}$ (correlated systematics)
and $\sigma_{c}$ (normalization uncertainties) for $p_T$ larger than
$6.5$ GeV/$c$, and $y_i^{th}$ is the calculated results at each
$p_T$ point. Shown in Fig.~\ref{fig:raa} (b) as two examples are the
$\chi^2$ distributions as functions of $\hat{q}_0 \tau_0$
($\tau_{0}$=0.6 fm/$c$) from the fits with the NLO pQCD calculation
with (1+1)d Bjorken model for the bulk medium evolution. The best
fit value is $\hat{q}_0\tau_0=0.36 \mbox{GeV}^2$ and 0.39
$\mbox{GeV}^2$
 for Woods-Saxon and hard-sphere nuclear distribution function, respectively.
In the following calculations, we will always use the Woods-Saxon nuclear distribution function.

Listed in Table~\ref{table1} are the best fit values of $\hat{q}_0
\tau_0$ and the corresponding $\chi^{2}$ for jet quenching in
various models for the bulk medium evolution. We will discuss and
compare them in detail in the  remainder of this section.

\begin{table}[htbp]
\begin{center}
\begin{tabular}{l|ccccc}
\hline \hline Model & flow
&PT &$\hat{q}_0\tau_0 (\mbox{GeV}^2)$ & $\chi^{2}$ \\
\hline (1+1)d Bjorken [W-S] &  &
  & 0.36 & 78 \\
\hline (1+3)d Hydro &   &   & 0.42 & 116 \\
& $\surd$ & & 0.48 & 80 \\
& $\surd$ &$\surd$ & 0.54 & 106 \\
($\hat{q}_0^h=0$) & $\surd$ &$\surd$ & 0.78 & 71 \\
\hline BAMPS &  &  & 0.72 & 95 \\
& $\surd$ &  & 0.84 & 84 \\
& $\surd$ & $\surd$ & 0.96 & 96 \\
($\hat{q}_0^h=0$) & $\surd$ & $\surd$ & 1.17 & 86 \\
\hline \hline
\end{tabular}
\caption{Initial jet transport parameter $\hat{q}_0\tau_0$ and the
corresponding $\chi^{2}$. The check sign ($\surd$) denotes the
inclusion of flow or phase transition (PT) in the calculations,
respectively. The initial $\tau_0$ is 0.6 fm/c for hydrodynamic and
0.3 fm/$c$ for BAMPS model.} \label{table1}
\end{center}
\end{table}

\subsection{(1+1)d Bjorken versus (1+3)d Ideal Hydrodynamical Model}\label{sec:d-b}

\begin{figure*}[t]
\begin{center}
\includegraphics[width=140mm]{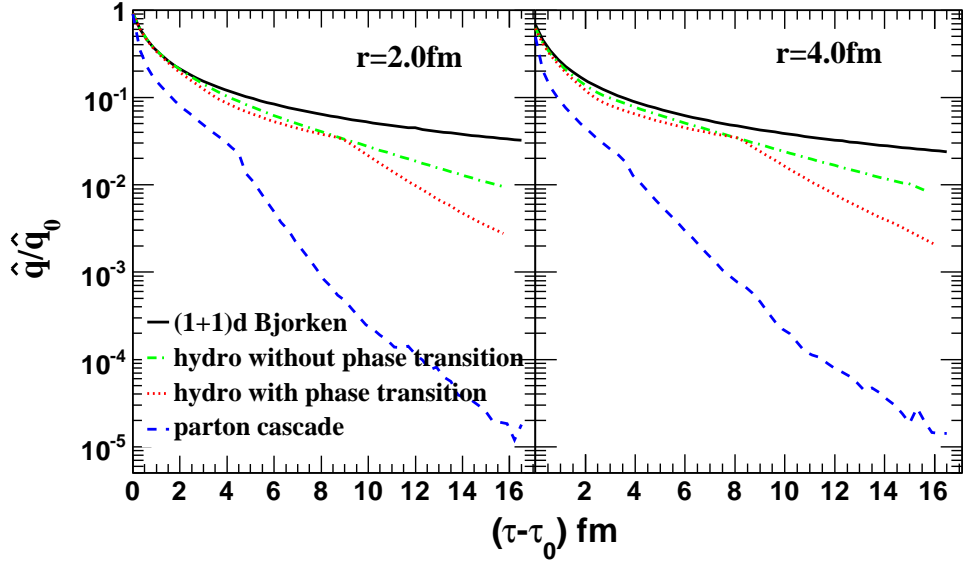}
\caption{(color online). Time evolution of the scaled jet transport
parameter at position of radius of $2$ fm (left panel) and $4$ fm
(right panel) for various models of the bulk matter evolution. The
flow effect in Eq.(\ref{q-hat-flow}) is not included.}
\label{fig:qhat}
\end{center}
\end{figure*}

We first compare the results from (1+1)d Bjorken and (1+3)d ideal
hydrodynamic model with different options and study the effects of
the transverse expansion, radial flow and
contributions from jet quenching in the hadronic phase. Comparing
the (1+1)d Bjorken and (1+3)d ideal hydrodynamical evolution without
radial flow and phase transition it is clear that the transverse
expansion cools the system considerably faster in the later stage of
the evolution. This faster cooling due to the transverse expansion
leads to a faster decrease of the jet transport parameter in the
(1+3)d hydrodynamic model than in the (1+1)d Bjorken model as shown
in Fig.~\ref{fig:qhat} by the dot-dashed [(1+3)d hydrodynamics] and
solid lines [(1+1)d Bjorken hydrodynamics]. To compensate for this
faster cooling, one has to increase the value of the initial jet
transport parameter $\hat q_{0}$ by 17\% in the case of (1+3)d
hydrodynamic evolution relative to the case of (1+1)d Bjorken
expansion in order to fit the measured suppression factor
$R_{AA}(p_{T})$ for neutral pions in the central $Au+Au$ collisions
at RHIC.

\begin{figure*}[t]
\begin{center}
\includegraphics[width=140mm]{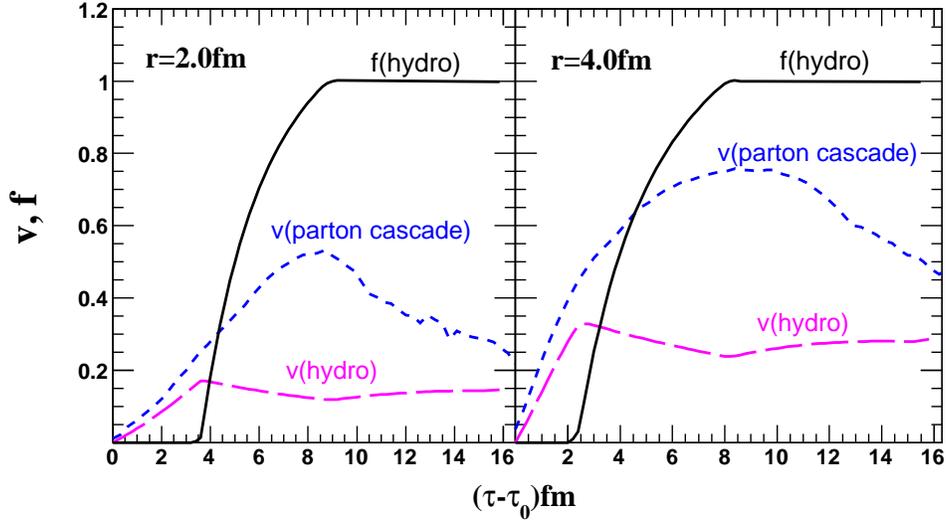}
\caption{(color online). The radial flow velocity in the (1+3)d
hydrodynamical model and the parton cascade at position of radius of
$2$ fm (left panel) and $4$ fm (right panel). The solid curves show
the fraction of the hadron phase in the hydrodynamical model.}
\label{fig:flow}
\end{center}
\end{figure*}

Inclusion of the radial flow in the calculation of the jet transport parameter as in Eq.~(\ref{q-hat-flow}) will also
change the effective jet quenching throughout the evolution of the bulk medium. Shown in Fig.~\ref{fig:flow}
are the time evolution of the radial flow velocity in (1+3)d hydrodynamics (dashed line).  The radial
flow velocity in the center of the dense medium (left panel), where jet quenching is the strongest, remains
small, at about 0.2, throughout hydrodynamic expansion.  In the same figure, we also plot the fraction of the
hadron phase (solid lines) as a function of time in the (1+3)d hydrodynamics. The period
between $f=0$ and $f=1$ indicate the duration of the phase transition or the mixed phase.
We notice that at the beginning of the mixed phase the radial flow
velocity starts to saturate and even decrease a little. This is mainly due to the mass effect.
The energy, approximately $m_T v$, where $m_T$ is the transverse mass,
should be the same during the phase transition. Therefore, the radial flow velocity will decrease
as particles become massive during the phase transition. Such decrease will be partially compensated
by the continued collective expansion which tend to increase the flow velocity. As the phase transition
is completed, the radial flow velocity should increase again by further transverse expansion in the
hadronic phase and  the resonances decays.

The radial flow will reduce the effective jet transport parameter
$\hat q$ if jets propagate along with it while increase the
effective value of $\hat q$ if jets propagate against it. After
averaging over the direction and initial position of jet production,
the effective jet transport parameter is merely reduced by less than
14\% \cite{flow1}. As shown in Table \ref{table1}, this leads to an
increase of $\hat q_{0}$ about 14\% when one includes  the effect of
radial flow in the (1+3)d hydrodynamic evolution in fitting
experimental data on $R_{AA}(p_{T})$.

In both (1+1)d Bjorken and two options of the (1+3)d hydrodynamic
expansion as discussed above, we have neglected the difference of
the jet transport parameters in QGP and hadronic phases. These
options are labelled as no phase-transition (PT) by setting $f=0$ in
Eq.~(\ref{q-hat-qgph}) and using
$\rho^{QGP}=a_{1}[(\epsilon-B)/b_{1}]^{3/4}$ in Eq.~(\ref{rhoqgp})
throughout the evolution even in the mixed and hadronic phase.

In the realistic scenario,  jet transport
parameter $\hat q_{h}$ in the hadronic phase should be different
from that in the QGP phase and we assume that $\hat q_{h}$ is
proportional to $\hat q_{N}$ in the cold nuclear medium and the
hadron number density as in Eq.~(\ref{eq:qhath}). With this
assumption, there is a discontinuity in $\hat q$ as a function of
time (or temperature as in Fig.~\ref{fig:qhat-t3}) at the end of the mixed phase,
as show by the dotted lines in Fig.~\ref{fig:qhat}, and the jet transport parameter
in the hadronic phase in the later stage of the bulk medium
evolution will be smaller than the scenario of no PT in the (1+3)d
hydrodynamics. Therefore, one needs a larger value of the initial
jet transport parameter $\hat q_{0}$, about 13\%, to account for the
measured suppression factor $R_{AA}(p_{T})$ as compared to an
evolving bulk medium with the same jet transport parameter
throughout difference phases.

To determine the contribution of  jet quenching in the hadronic
phase, we set $\hat{q}^h_0=0$ in Eq.~(\ref{q-hat-qgph}), assuming
parton energy loss only happens in the QGP phase. In this case, the
extracted $\hat{q}_0$ will be about  44\% larger (see
Table~\ref{table1}). This implies that the hadronic phase
contributes to about 44\% of the total suppression of the high
$p_{T}$ hadron spectra. Such a large contribution is due to the
lifetime of the mixed and hadronic phase until the kinetic freeze-out which is
longer than the lifetime of the QGP phase, though the hadron density
is much smaller than the initial parton density in the QGP phase. A
better understanding of the hadronic contribution to the jet
quenching is, therefore, important for an accurate extraction of the
jet transport parameter in the initial phase of strongly interacting
QGP.

\subsection{(1+3)d Parton Cascade versus (1+3)d Ideal Hydrodynamical Model}
\label{sec:d-c}

In a parton cascade model such as BAMPS, the system takes times to
reach thermal or partial thermal equilibrium after the initial
parton production time, $\tau_{0}=0.3$ fm/$c$. During this period of
parton equilibration, there is entropy production accompanied by
rapid expansion. The net effect is a much faster decreasing of the
effective temperature \cite{Xu:2004mz,Xu:2007aa,Biro:1993qt} and the
parton density during the early stage of the bulk medium evolution.
The corresponding jet transport parameter in the equilibrating
gluonic matter in BAMPS calculation therefore decreases much faster
than that in both (1+1)d and (1+3)d ideal hydrodynamic models in the
early stage of evolution as shown by the dashed lines in
Fig.~\ref{fig:qhat}. It is smaller by more than a factor of 2 before
the hadronization. This leads to about 77\% increase in the initial
value of $\hat q_{0}$ that is needed to account for the measured
suppression of hadron spectra (see Table~\ref{table1}). Note that in
our calculations jet quenching starts at the thermalization
$\tau_0=0.6$ fm/$c$ in the ideal hydrodynamic model. However, it
starts at $\tau_0=0.3$ fm/$c$ in the BAMPS model immediately after
the initial gluons are produced and before thermal or partial
thermal equilibrium is reached. Such jet medium interaction and
parton energy loss during the thermalization stage in the BAMPS
model will contribute to some additional, though a small fraction,
suppression of the final hadron spectra.

In BAMPS, the hadronization of the gluonic  matter happens when the
local gluon energy density is below $\epsilon_{c}=0.6 \mbox{GeV/
fm}^3$ and parton-hadron duality is used for the phase-space density
of hadrons, which are assumed to freeze out immediately. Such a
sudden hadronization and freeze-out scenario implies the change of the hadron
fraction $f$ from 0 to 1 at this critical density in the evaluation
of the jet transport parameter in Eq.~(\ref{q-hat-qgph}). The lack
of the mixed phase in BAMPS, which lasts for about $5-6$ fm/$c$ in
the (1+3)d ideal hydrodynamic evolution as indicated by the duration
for the hadron fraction to reach $f=1$ in Fig.~\ref{fig:flow}, leads
to smaller value of the jet transport parameter  after the
hadronization. Comparing the extracted values of $\hat q_{0}$ from
the phenomenological fit with and without ($\hat q_{h}=0$) hadronic
phase as shown in Table~\ref{table1}, the hadronic interaction in
BAMPS model contributes to about 22\% of the jet quenching effect,
which is about a factor 2 smaller than that in the (1+3)d ideal
hydrodynamical model.

The large viscous effect in the parton cascade model
\cite{Xu:2007jv} generally  leads to larger radial flow velocity
than that generated by an ideal hydrodynamic expansion as shown in
Fig.~\ref{fig:flow}. The free-streaming of hadrons immediately after
the hadronization also leads to the further increase of the flow
velocity. Such larger radial flow velocity in the BAMPS model
reduces the effective jet quenching parameter and therefore leads an
increase of about 16\% in the extracted initial $\hat q_{0}$ .

The decrease of the radial flow velocity at late times seen in Fig.~\ref{fig:flow}
is a numerical artifact, which comes from the chosen geometry of
spatial elements, where particles are collected to calculate the
radial flow velocity. Longitudinal length of each spatial element
corresponds to a space time rapidity window $|\eta| < 0.25$, while
the transverse length is fixed to be 0.3 fm. At late times the
longitudinal length is larger than the transverse one, and
particles with smaller transverse velocity (larger longitudinal
velocity) make up the main part of remaining particles in the spatial
elements. This is the reason for the observed decrease of the radial
flow velocity. If we choose the longitudinal length being equal as
the transverse one at the late times, we would see a continuous
increase of the flow velocity due to the free-streaming, before all
particles go out of the spatial element. This is, however, a
negligible effect on the extraction of $\hat q_{0}$, because the
hadron density is tiny at late times.

\section{Conclusions}\label{sec:e}

Using medium-modified fragmentation functions from the high-twist
approach to multiple parton scattering we have calculated single
inclusive hadron spectra in NLO perturbative QCD parton model. The
medium modification to the fragmentation function is very similar in
form to the DGLAP correction due to vacuum gluon bremsstrahlung,
except that the medium modified splitting function contains
information about the properties of the medium through a
path-integrated jet transport parameter $\hat q$. Therefore, the
medium modified fragmentation function will explicitly depend on the
space-time evolution of the bulk medium. We have focused our
attention on the dependence of the medium modified fragmentation
function on the space-time evolution of the bulk medium and their
influence on the suppression of the single hadron spectra at large
$p_{T}$.

We specifically considered the effects of transverse expansion,
radial flow, jet quenching in hadronic phase and non-equilibrium
effect with three different models for the bulk matter evolution in
central $Au+Au$ collisions at the RHIC energy: (1+1)d Bjorken
hydrodynamic model, (1+3)d ideal hydrodynamical model by Hirano
\cite{hirano1, hirano2} and (1+3)d BAMPS \cite{Xu:2004mz,Xu:2007aa}
parton cascade model. Since the jet transport parameter is
proportional to the particle gluon distribution density of the
medium, we have assumed that it will be proportional to the local
particle density in both QGP and hadronic phase during the bulk
matter evolution. The coefficient of the particle density
dependence, which is related to soft gluon distribution per particle,
will be different in the QGP and hadronic
phase. Assuming a similar density dependence of the jet transport
parameter in the hadronic phase as that in a large cold
nuclei, but rescaled by the local hadron density and the number of
constituent quarks, the only free parameter in the space-time profile is
the initial value of the jet transport parameter $\hat q_{0}$ at
initial time $\tau_{0}$ at the center of the QGP medium in central
heavy-ion collisions. We have extracted values of $\hat{q}_0$ at
$\tau_0$ in each model of bulk evolution with different options by
fitting the experimental data on the suppression factor
$R_{AA}(p_{T})$ for high $p_{T}$ neutral pions and studied the
uncertainties in the extracted value of $\hat q_{0}$.

By comparing the values of $\hat q_{0}$ extracted with different
options in models for the bulk evolution, we also found that the
transverse expansion leads to fast cooling of the hot medium and
therefore reduces jet quenching in the later stage of the evolution
by about 17\%. Radial flow also reduces the effective jet transport
parameter by 15\%. The phase transition from QGP to hadronic medium
further reduces the effective jet transport parameter in the
hadronic phase and leads to about 13\% reduction of jet quenching
effect. Within our model for the form of jet transport parameter due
to jet-hadron interaction, the overall contribution to jet quenching
from hadronic phase is about 22-44\%, depending on the model for
evolution of the hadronic phase.

By comparing (1+3)d ideal hydrodynamic and the BAMPS parton cascade
model, we found that the non-equilibrium evolution in the early
stage of parton cascade leads to faster decrease of the jet
transport parameter with time and therefore affects the overall jet
quenching the most by 77\%, part of which is caused by the lack of
mixed phase in the parton cascade model. Such uncertainties can be
further reduced by combining parton cascade and hadron cascade with
a model of of hadronization that can reproduce the EOS during the
phase transition as given by lattice QCD calculation.

\section*{Acknowledgements}

We would like to thank T. Hirano for providing numerical results
of (1+3)d ideal hydrodynamic evolution of the bulk matter in
heavy-ion collisions at RHIC and discussions. This work is supported
by the NSFC of China under Projects No. 10825523, No. 10635020, by
MOE of China under Projects No. IRT0624, by MOST of China under
Project No. 2008CB317106 and by MOE and SAFEA of China under Project
No. PITDU-B08033, and by the Director, Office of Energy Research,
Office of High Energy and Nuclear Physics, Divisions of Nuclear
Physics, of the U.S. Department of Energy under Contract No.
DE-AC02-05CH11231. The numerical calculations were performed at the
Center for Scientific Computing of Goethe University. This work was
financially supported by the Helmholtz International Center for FAIR
within the framework of the LOEWE program (Landes-Offensive zur
Entwicklung Wissenschaftlich-\"okonomischer Exzellenz) launched by
the State of Hesse, Germany. X.-N. Wang thanks the hospitality of
the Institut f\"ur Theoretische Physik, Johann Wolfgang
Goethe-Universit\"at and support by the ExtreMe Matter Institute
EMMI in the framework of the Helmholtz Alliance Program of the
Helmholtz Association (HA216/EMMI) during the completion of this
work.

\end{document}